 \documentclass[final,5p,times,twocolumn]{elsarticle}

\newcommand{\mumax}{\textsc{MuMax}\xspace}

\usepackage{color}
\usepackage{graphicx}
\usepackage{xspace}
\usepackage[utf8]{inputenc}
\usepackage{amssymb}
\usepackage{amsmath}

\begin{document}

\begin{frontmatter}

\title{\mumax: a new high-performance micromagnetic simulation tool}

\author[label1]{A. Vansteenkiste}
\ead{Arne.Vansteenkiste@ugent.be}
\author[label2]{B. Van de Wiele}

\address[label1]{Department of Solid State Sciences, Ghent University, Krijgslaan 281-S1, B9000 Gent, Belgium.}
\address[label2]{Department of Electrical Energy, Systems and Automation, Ghent University, Sint Pietersnieuwstraat 41, B-9000 Ghent, Belgium.}


\begin{abstract}

We present \mumax, a general-purpose micromagnetic simulation tool running on Graphical Processing Units (GPUs). \mumax is designed for high performance computations and specifically targets large simulations. In that case speedups of over a factor 100$\times$ can easily be obtained compared to the CPU-based OOMMF program developed at NIST. \mumax aims to be general and broadly applicable. It solves the classical Landau-Lifshitz equation taking into account the magnetostatic, exchange and anisotropy interactions, thermal effects and spin-transfer torque.  Periodic boundary conditions can optionally be imposed.  A spatial discretization using finite differences in 2 or 3 dimensions can be employed. \mumax is publicly available as open source software. It can thus be freely used and extended by community. Due to its high computational performance, \mumax should open up the possibility of running extensive simulations that would be nearly inaccessible with typical CPU-based simulators.
\end{abstract}

\begin{keyword}
micromagnetism \sep simulation \sep GPU
\PACS 75.78.Cd \sep 02.70.Bf
\end{keyword}

\end{frontmatter}

\section{Introduction}

Micromagnetic simulations are indispensable tools in the field of magnetism research. Hence, micromagnetic simulators like, e.g., OOMMF \cite{oommf}, magpar \cite{Scholz03} and Nmag \cite{nmag} are widely used. These tools solve the Landau-Lifshitz equation on regular CPU hardware. Due to the required fine spatial and temporal discretizations, such simulations can be very time consuming. Limited computational resources therefore often limit the full capabilities of the otherwise successful micromagnetic approach.

There is currently a growing interest  in running numerical calculations on graphical processing units (GPUs) instead of CPUs.  Although originally intended for purely graphical purposes, GPUs turn out  to be well-suited for high-performance, general-purpose calculations. Even relatively cheap GPUs can perform an enormous amount of calculations in parallel. E.g., the nVIDIA GTX580 GPU used for this work costs less than \$\,500 and delivers 1.5 trillion floating-point operations (Flops) per second, about 2 orders of magnitude more than a typical CPU.  

However, in order to employ this huge numerical power programs need to be written specifically for GPU hardware, using the programming languages and tools provided by the GPU manufacturer, and the code also needs to handle many hardware-specific technicalities. Additionally,  the used algorithms need to be expressed in a highly parallel manner, which is not always easily possible.

Other groups have already implemented micromagnetic simulations on GPU hardware and report considerable speedups compared to a CPU-only implementation \cite{Kakay2010, Li2010, Selke2011}. At the time of writing, however, none of these implementations is freely available. \mumax, on the other hand, is available as open source software and can be readily used by anyone. Its performance also compares favorably to these other implementations.

\section{Methods}

Since the micromagnetic theory describes the magnetization as a continuum field $\mathbf{M}(\mathbf{r}, t)$, the considered magnetic sample is discretized in cuboidal finite difference (FD) cells with a uniform magnetization. The time evolution of the magnetization in each cell is given by the Landau-Lifshitz equation

\begin{equation}
\begin{split}
\frac{\partial\mathbf{M}(\mathbf{r}, t)}{\partial t} =
&- \frac{\gamma}{1+\alpha^2}\mathbf{M}(\mathbf{r}, t)\times \mathbf{H}_{eff}(\mathbf{r}, t)\\
&- \frac{\alpha \gamma}{M_s(1+\alpha^2)}\mathbf{M}(\mathbf{r}, t)\times\left(\mathbf{M}(\mathbf{r}, t)\times\mathbf{H}_{eff}(\mathbf{r}, t)\right).\label{LLequation}
\end{split}
\end{equation}
Here, $M_s$ is the saturation magnetization, $\gamma$ the gyromagnetic ratio and $\alpha$ the damping parameter.  The continuum effective field $\mathbf{H}_{eff}$ has several contributions that depend on the magnetization, the externally applied field and the material parameters of the considered sample. When timestepping equation (\ref{LLequation}) the effective field is evaluated several times per time step.  Hence, the efficiency of micromagnetic software depends on the efficient evaluation of the different effective field terms at the one hand and the application of efficient time stepping schemes on the other hand.  \mumax combines both with the huge computational power of GPU hardware.

\subsection{Effective field terms}
In the present version of \mumax, the effective field can have 5 different contributions: the magnetostatic field, the exchange field, the applied field, the anisotropy field and the thermal field.  In what follows we present these terms and comment on their optimized implementation.

\subsubsection{Magnetostatic field}
The magnetostatic field $\mathbf{H}_{ms}$ accounts for the long-range interaction throughout the complete sample
\begin{equation}
\mathbf{H}_{ms}(\mathbf{r})
= -\frac{1}{4\pi}\int_V \nabla\nabla\frac{1}{|\mathbf{r}-\mathbf{r}'|}\cdot\mathbf{M}(\mathbf{r}')
\,\mathrm{d}\mathbf{r}'.
\label{Hms}
\end{equation}
Since the magnetostatic field in one FD cell depends on the magnetization in all other FD cells, the calculation of $\mathbf{H}_{ms}$ is the most time-consuming part of a micromagnetic simulation. The chosen method for this calculation is thus decisive for the performance of the simulator. Therefore, we opted for a fast Fourier transform (FFT) based method. In this case, the convolution structure of (\ref{Hms}) is exploited.  By applying the convolution theorem, the convolution is accelerated by first Fourier transforming the magnetization, then multiplying this result with the Fourier-transform of the convolution kernel and finally inverse transforming this product to obtain the magnetostatic field. The overall complexity of this method is $\mathcal{O}(N \log N)$, as it is dominated by the FFTs.

Methods with even lower complexity exist as well. The fast multipole method, e.g., only has complexity $\mathcal{O}(N)$, but with such a large pre-factor that in most cases the FFT method remains much faster \cite{VandeWiele2008}.

A consequence of the FFT method is that the magnetic moments must lie on a regular grid. This means that a finite difference (FD) spatial discretization has to be used: space is divided into equal cuboid cells. This method is thus most suited for rectangular geometries. Other shapes have to be approximated in a staircase-like fashion. However, thanks to the speedup offered by \mumax's, smaller cells may be used to improve this without excessive performance penalties. 

The possibility of adding periodic boundary conditions in one or more directions is also included in the software.  This is done by adding a sufficiently large number of periodic images to the convolution kernel.  The application of periodic boundary conditions has a positive influence on the computational time since the magnetization data does not need to be zero padded in the periodic directions, which roughly halves the time spend on FFTs for every periodic direction.

\subsubsection{Exchange field}

The exchange interaction contributes to the effective field in the form of a laplacian of the magnetization
\begin{equation}
\mathbf{H}_{exch} = \frac{2A}{\mu_0 M_s}\nabla^2\mathbf{m},\label{Hexch}
\end{equation}
with $A$ the exchange stiffness.  In discretized form, this can be expressed as a linear combination of  the magnetization of a cell and a number of its neighbors. \mumax uses a 6-neighbor scheme, similar to \cite{Donahue2004}. In the case of 2D simulations (only one FD cell in the z-direction), this method automatically reduces to a 4-neighbor scheme.

The exchange field calculation is included in the magnetostatic field routines by simply adding the kernel describing the exchange interaction to the magnetostatic kernel. In this way, the exchange calculation is essentially free, as only one joint convolution product is needed to simultaneously evaluate both the magnetostatic and exchange fields.  Moreover, by introducing the exchange contribution in the magnetostatic field kernel periodic boundary conditions are directly accounted for if applicable.

\subsubsection{Other effective field terms}
Next to the above mentioned interaction terms and the applied field contribution, \mumax provides the ability to include magnetocrystalline anisotropy. Currently, uniaxial and cubic anisotropy are available. The considered anisotropy energies are
\begin{equation}
\phi_{ani} = K_u\sin^2\theta
\end{equation}
and
\begin{equation}
\begin{split}
\phi_{ani}(\mathbf{r}) &= K_1\left[
  \alpha_1^2(\mathbf{r})\alpha_2^2(\mathbf{r})
+ \alpha_2^2(\mathbf{r})\alpha_3^2(\mathbf{r})
+ \alpha_1^2(\mathbf{r})\alpha_3^2(\mathbf{r})\right]\\
&+ K_2\left[\alpha_1^2(\mathbf{r})\alpha_2^2(\mathbf{r})\alpha_3^2(\mathbf{r})\right]\label{phi_ani}
\end{split}
\end{equation}
for uniaxial and cubical anisotropy respectively.  Here, $K_u$ and $(K_1, K_2)$ are the uniaxial and cubical anisotropy constants, $\theta$ is the angle between the local magnetization and uniaxial anisotropy axis and $\alpha_i$ ($i=1,2,3$) are the direction cosines between the local magnetization and the cubic easy magnetization axes.

Furthermore, thermal effects are included by means of a fluctuating thermal field
\begin{equation}
\mathbf{H}_{th} = \boldsymbol{\eta}(\mathbf{r},t)\sqrt{\frac{2\alpha k_B T}{\gamma\mu_0M_s V\delta t}}\label{Hth}
\end{equation}
which is added to the effective field $\mathbf{H}_{eff}$ according to \cite{Brown63}.  In (\ref{Hth}), $k_B$ is the Boltzmann constant, $V$ is the volume of a FD cell, $\delta t$ is the used time step and $\boldsymbol{\eta}(\mathbf{r},t)$ is a stochastic vector whose components are Gaussian random numbers, uncorrelated in space and time with zero mean value and dispersion 1. 

\subsubsection{Spin-transfer torque}
The spin-transfer torque interaction describes the influence of electrical currents on the local magnetization. Possible applications are spin-transfer torque random access memory \cite{bohlens08} and racetrack memory \cite{Parkin2008}. \mumax incorporates the spin-transfer torque description developed by Berger \cite{Berger1996}, refined by Zhang and Li \cite{Zhang2004}
\begin{equation}
\begin{split}
\frac{\partial \mathbf{M}}{\partial t} = &-\frac{\gamma}{1+ \alpha^2}\mathbf{M}\times \mathbf{H}_{eff} \\
&- \frac{\alpha\gamma}{M_s(1+\alpha^2)}\mathbf{M}\times(\mathbf{M}\times \mathbf{H}_{eff})\\
&- \frac{b_j}{M_s^2(1+\alpha^2)}\mathbf{M}\times\left(\mathbf{M}\times (\mathbf{j}\cdot\nabla)\mathbf{M}\right)\\
&- \frac{b_j}{M_s(1+\alpha^2)}(\xi-\alpha) \mathbf{M}\times (\mathbf{j}\cdot\nabla)\mathbf{M}.\label{STT}
\end{split}
\end{equation}
Here, $\xi$ is the degree of non-adiabicity and $b_j$ is the coupling constant between the current density $\mathbf{j}$ and the magnetization
\begin{equation}
b_j = \frac{P \mu_B}{eM_s(1+\xi^2)},
\end{equation}
with $P$ the polarization of the current density,  $\mu_B$ the Bohr magneton and $e$ the electron charge.

\subsection {Time integration schemes}

\mumax provides a range of Runge-Kutta (RK) methods to integrate the Landau-Lifshitz equation. Currently the user can select between the following options:
\begin{itemize}
	\item RK1: Euler's method
	\item RK2: Heun's method
	\item RK12: Heun-Euler (adaptive step)
	\item RK3: Kutta's method
	\item RK23: Bogacki–Shampine (adaptive step)
	\item RK4: Classical Runge-Kutta method
	\item RKCK: Cash-Karp (adaptive step)
	\item RKDP: Dormand–Prince (adaptive step)
\end{itemize}

The adaptive step methods adjust the time step based on a maximum tolerable error per integration step that can be set by the user. The other methods can use either a fixed time step or a fixed maximum variation of $\textbf{m}$ per step. Depending on the needs of the simulation, a very accurate but relatively slow high-order solver (e.g. RKDP) or a less accurate but fast solver (e.g. RK12) can be chosen. Additionally, \mumax incorporates the semi-analytical methods described in \cite{VandeWiele2007}. These methods are specifically tailored to the Landau-Lifshitz equation.

\section{GPU-optimized implementation}

Since various CPU based micromagnetic tools ---well suited for relatively small micromagnetic problems--- are already available, we mainly concentrated on optimizing \mumax for running very large simulations on GPUs. Nevertheless, the code can also run in CPU-mode, with multi-threading modalities enabled. In this way one can get familiar with the capabilities of \mumax before a high-end GPU has to be purchased.  

The GPU-specific parts of \mumax have been developed using nVIDIA's CUDA platform. The low-level, performance-critical functions that have to interact directly with the GPU are written in C/C++. Counterparts of these functions for the CPU  are implemented as well and use FFTW \cite{fftw} and multi threading. The high-level parts of \mumax are implemented in "safe" languages including Java, Go and Python. This part is independent of the underlying GPU/CPU hardware. In what follows we will only elaborate on the GPU-optimized implementation of the low-level functions. 

\subsection{General precautions}

A high-end GPU has its own dedicated memory with a high bandwidth (typically a few hundred GB/s) which enables fast reads and writes on the GPU itself. Communication with the CPU on the other hand is much slower since this takes place over a PCI express bus with a much lower bandwidth (typically a few GB/s). Therefore, our implementation keeps as much data as possible in the dedicated GPU memory, avoiding CPU-GPU communication.  The only large data transfers occure at initialization time and when output is saved to disk. The CPU thus only instructs the GPU which subroutines to launch. Hence, the GPU handles all the major computational jobs.

On the GPU, an enormous number of threads can run in parallel, each performing a small part of the computations. E.g., the GTX580 GPU used for this work has 512 computational cores grouped in 16 multiprocessors, resulting in total number of 16384 available parallel threads. However, this huge parallel power is only optimally exploited when the code is adapted to the specific GPU architecture. E.g.: threads on the same multiprocessor ("thread \emph{warps}") should only access the GPU memory in a coalesced way and should ideally perform the same instructions. When coalesced memory access in not possible, the so-called \emph{shared memory} should be used instead of the global GPU memory. This memory is faster and has better random-access properties but is very scarce. Our implementation takes into account all these technicalities, resulting in a very high performance. 

\subsection{GPU-optimization of the convolution product}

Generally, most computational time goes to the evaluation of the convolution product defined by the magnetostatic field.  When using fast Fourier transforms (FFTs), the computations enhance three different stages: (i) forward Fourier transforming the magnetization data that is zero padded in the non-periodic directions, (ii) point-by-point multiplying the obtained data with the Fourier transformed magnetostatic field kernel, (iii) inverse Fourier transforming the resulting magnetostatic field data. The carefull implementation of these three stages determines the efficiency of the convolution product and, more general, of the micromagnetic code.

In the first place, the efficiency of this convolution process is safeguarded by ensuring that the matrices defined by the magnetostatic field kernel are completely symmetrical.  Consequently, the Fourier transformed kernel data is purely real.  The absence of the imaginary part leads to smaller memory requirements as well as a much faster evaluation of the point-by-point multiplications -- step (ii).
 
Furthermore, our GPU implementation of the fast Fourier transforms, which internally uses the CUDA "CUFFT" library, is specifically optimized for micromagnetic applications. The general 3D real-to-complex Fourier transform (and its inverse) available in the CUFFT library is replaced by a more efficient implementation in which the set of 1D transforms in the different directions are performed separately. This way, Fourier transforms on arrays containing only zeros resulting from the zero padding are avoided.  In each dimension, the set of 1D Fourier transforms are performed on contiguous data points resulting in the coalesced reading and writing of the data.  As a drawback, the transposition of the data between a set of Fourier transforms in one and another dimension is needed.  

In a straight forward implementation of the required matrix transposes, the read and write instructions can not be both performed in a coalesced way since either the input data or the transposed data is not contiguous in global memory.  Therefore, the input data is divided in blocks and copied to shared memory assigned to a predefined number of GPU threads.  There, the data block is transposed and copied in a coalesced way back to the global GPU memory space.  By inventively using the large number of zero arrays --in the non-periodic case-- this transpose process can be done without (for 2D) or with only limited (3D) extra memory requirements.  The different sets of Fourier transforms in this approach are performed using the 1D FFT routines available in the CUFFT library. This implementation outperforms the general 3D real-to-complex Fourier transform available in the CUFFT library while the built-in 2D real-to-complex is only faster for small dimensions (for square geometries: smaller than 512x512 FD cells).  This approach ensures the efficient evaluation of steps (i) and (iii) of the convolution.

\subsection{Floating point precision}

GPUs are in general better suited for single-precision than double-precision arithmetic. Double-precision performance is not only much slower due to the smaller number of arithmetic units, but also requires twice the amount of memory.  Since GPU's typically have limited memory and FFT methods are relatively memory-intensive, we opted to uses single-precision exclusively.

While, e.g., the finite element method used by Kakay et al. relies heavily on double precision to obtain an accurate solution  \cite{Kakay2010}, our implementation is designed to remain accurate even at single precission. First, all quantities are internally stored in units that are well adapted to the problem. More specifically, we choose units so that $\mu_0 = \gamma_0 = M_s = A = 1$. This avoids that any other quantity in the simulation becomes exceptionally large or small ---which could cause a loss of precision due to saturation errors. The conversion to and from internal units is performed transparently to the user. Secondly, we avoid numerically instable operations like, e.g., subtracting nearly equal numbers. This avoids that small rounding of errors get amplified. Finally, and most importantly, we restrict the size of the FFTs to numbers where the CUFFT implementation is most accurate: $2^n \times \{1,3,5 \mathrm{\ or\ } 7\}$. Hence sometimes a slightly larger number of FD cells than strictly necessary is used to meet this requirement. Fortunately this has no adverse effect on the performance since CUFFT FFTs with these sizes also happen to be exceptionally fast (see below).  In this way, the combined error introduced by the forward+inverse FFT was found to be only of the order of $\mathcal{O}(10^{-6})$, as opposed to a typical error of $\mathcal{O}(10^{-4})$ for other FFT sizes (estimated from the error on transforming random data back and forth). Thanks to these precautions we believe that our implementation should be sufficiently accurate for most practical applications. Indeed, the uncertainty on material parameters alone is usually much larger than the FFT error of $10^{-6}$.


\section{Validation}

In order to validate our software, we tested the reliability of the code by simulating several standard problems.  These standard problems are constructed such that all different contributions in the considered test case influence the magnetization processes significantly.  A correct simulation of standard problems can be considered as the best possible indication of the validity of the developed software. In what follows, we consider standard problems constructed for testing static simulations, dynamic simulations and dynamic simulations incorporating spin-transfer torque.

\subsection{static standard problem}\label{prob2}

The $\mu$MAG standard problem \#2 \cite{mumag} aims at testing quasi static simulations.  A cuboid with dimensions $5d\times d \times 0.1d$ is considered.  Since only magnetostatic and exchange interactions are included, the resulting static properties only depend on the scaled parameter $d/l_{ex}$, with $l_{ex}$ the exchange length.  The starting configuration is saturation along the $[1,1,1]$ axis, which is relaxed to the remanent state. 
This was done by solving the Landau-Lifshitz equation with a high damping parameter $\alpha=1$.

\begin{figure}
\begin{center}
\includegraphics[width = \columnwidth]{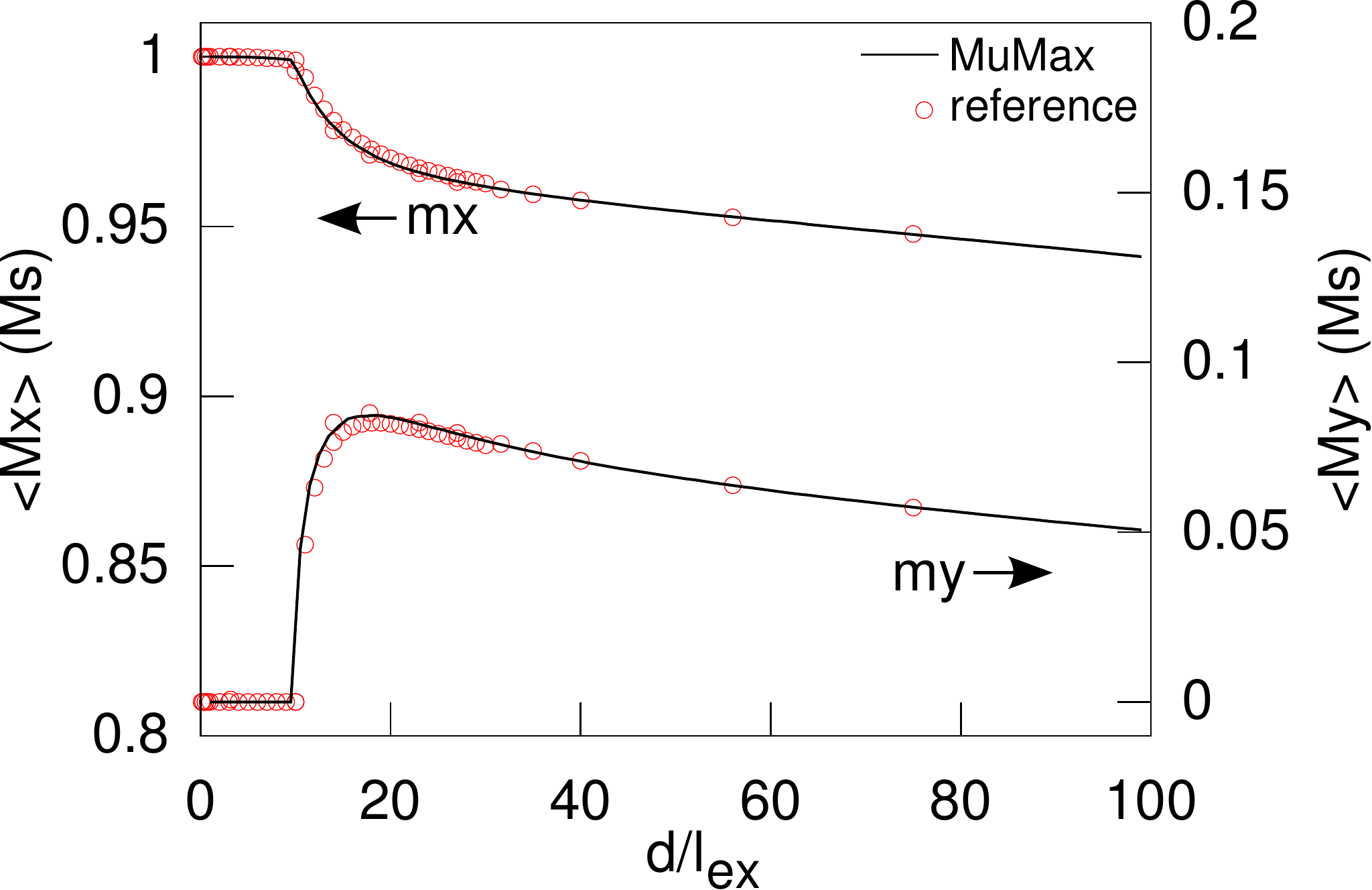}
\end{center}
\caption{Standard problem \#2. Remanent magnetization along the $x$-axis (left axis) and along the $y$-axis (right axis) as a function of the scaling parameter $d$.  The full line represents the simulation results from \mumax, while the circles represent simulation points obtained from McMichael et al. \cite{McMichael1999} and from Donahue et al.\cite{Donahue2000} \label{stdprobl2}}
\end{figure} 
 
The number of FD cells was chosen depending on the size of nanostructure, making sure the cell size remained below the exchange length. For $d/l_{ex}\leq 10$, single-domain states with nearly full saturation along the long axis were found, while for large geometries an S-state occured. The \mumax simulations considered 200 values for $d/l_{ex}$.  On the GPU a total simulation time of 3'21" was needed to complete the 200 individual simulations, compared to 34'30" on the CPU. The GPU speedup is here limited by the relatively small simulation sizes (cfr. Fig. \ref{performance}).

Figure \ref{stdprobl2} shows the remanent magnetization in function of the ratio $d/l_{ex}$. The values obtained with \mumax coincide well with those of other authors \cite{McMichael1999,Donahue2000}, validating \mumax for static micromagnetic problems.  
\subsection{dynamic standard problem }
\begin{figure}
\begin{center}
\includegraphics[width = \columnwidth]{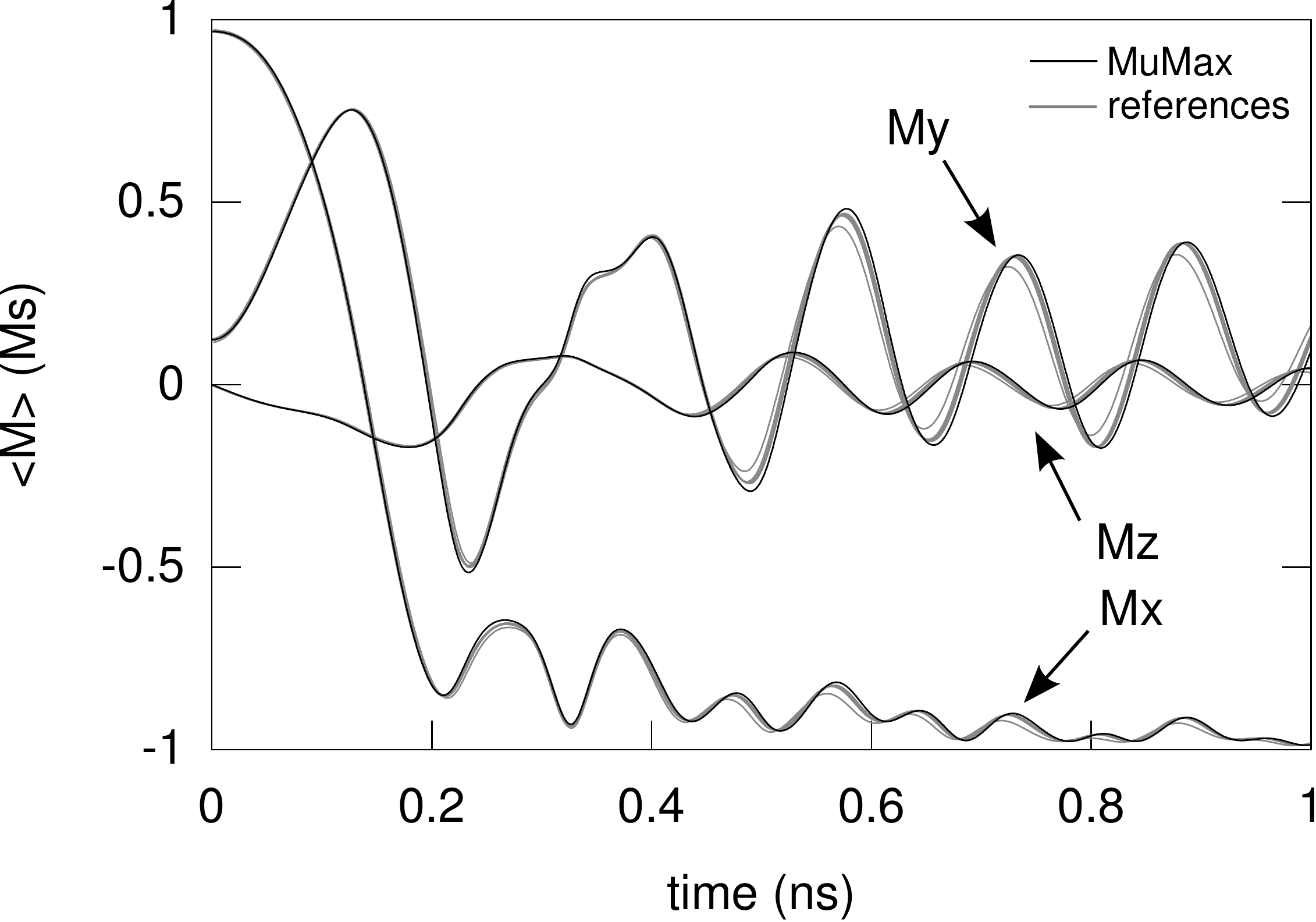}\\ \vspace{5 mm}
\includegraphics[width = \columnwidth]{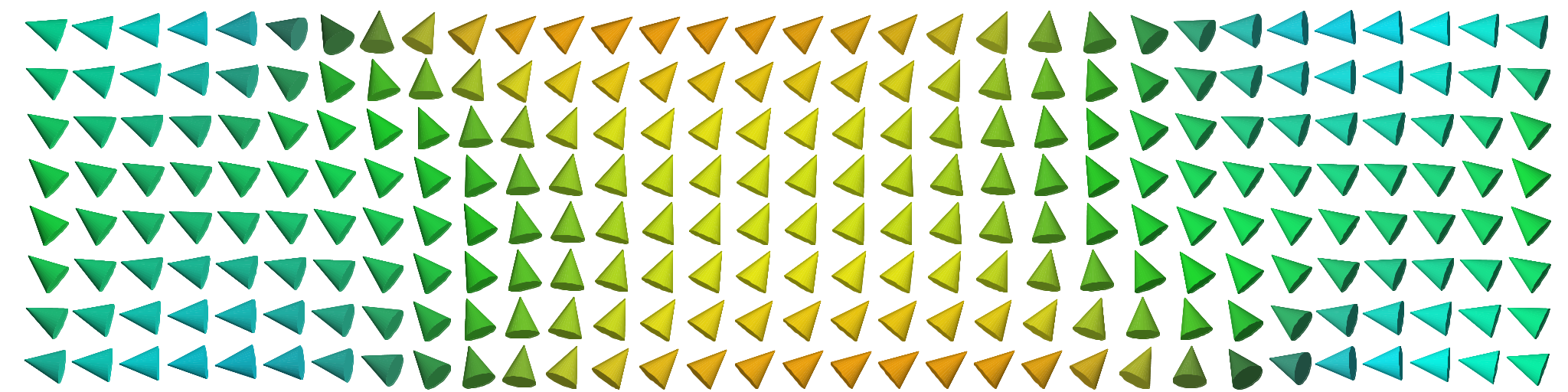}
\end{center}
\caption{(top) Time evolution of the average magnetization during the reversal considered in ${\mu}$Mag standard problem \#4, field1.  The results obtained with \mumax (black) lie well within the spread of the reference solutions (grey), taken from \cite{mumag}. (bottom) Magnetization configuration when $<M_x>$ crosses the zero magnetization for the first time. \label{stdprobl4_1}}
\end{figure}

\begin{figure}
\begin{center}
\includegraphics[width = \columnwidth]{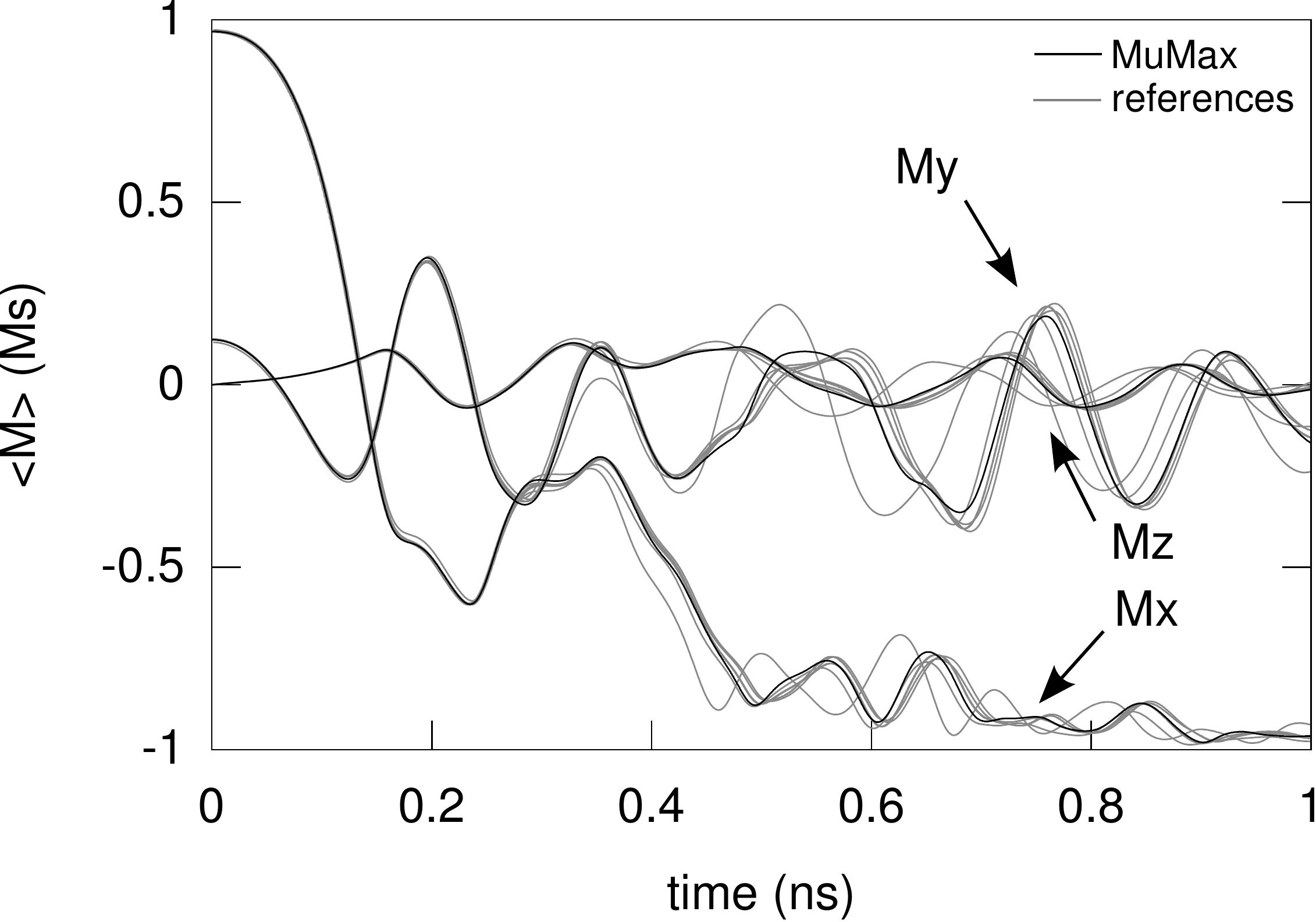}\\ \vspace{5 mm}
\includegraphics[width = \columnwidth]{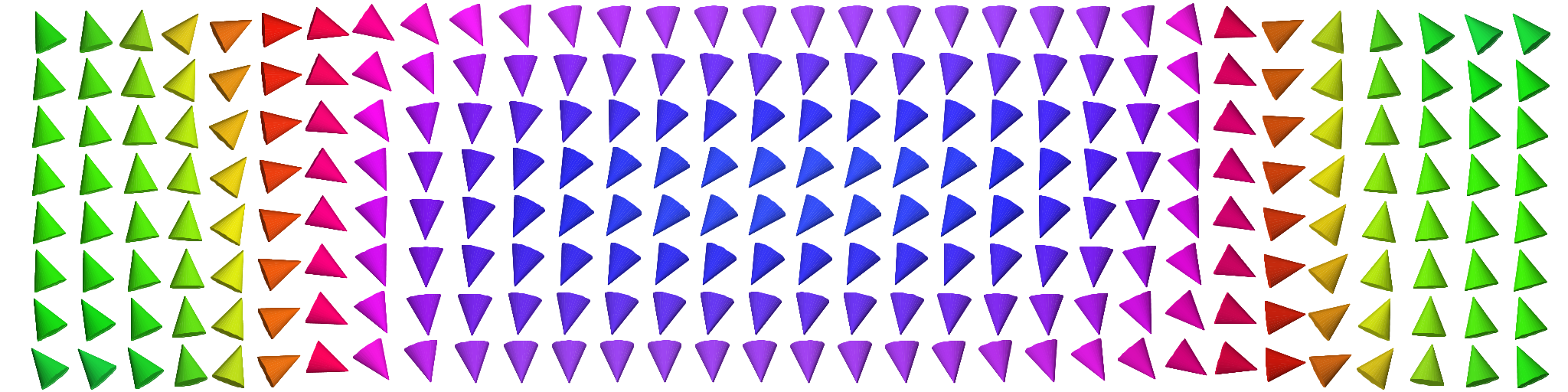}
\end{center}
\caption{(top) Time evolution of the average magnetization during the reversal considered in $\mu$Mag standard problem \#4, field2. This field was chosen to cause a bifurcation point to make the different solutions diverge. (bottom) Magnetization configuration when $<M_x>$ crosses the zero magnetization point for the first time.  \label{stdprobl4_2}}
\end{figure}

The $\mu$MAG standard problem \#4 \cite{mumag} aims at testing the description of the dynamic magnetization processes  by considering the magnetization reversal in a thin film with dimensions 500\,nm\,$\times$\,125\,nm\,$\times$\,3\,nm.  Starting from an initial equilibrium S-state, two different fields are applied.  In this problem only the exchange and magnetostatic interactions are considered (exchange stiffness $A=1.3\times10^{-11}$\,J/m, saturation magnetization $M_s=8.0\times10^5$\,Am$^{-1}$).  When relaxing to the initial equilibrium S-state, a damping constant equal to 1 is used while during the reversal itself a damping constant of 0.02 is applied, according to the problem definition.
As proposed in the standard problem, we show in Figs. \ref{stdprobl4_1} and \ref{stdprobl4_2} the evolution of the average magnetization components together with the reference magnetization configuration at the time point when $<M_x>$ crosses zero for the first time, for field 1 ($\mu_0H_x$=-24.6\,mT, $\mu_0H_y$= 4.3\,mT, $\mu_0H_z$= 0.0\,mT) and field 2 ($\mu_0H_x$=-35.5\,mT, $\mu_0H_y$= -6.3\,mT, $\mu_0H_z$= 0.0\,mT).  A discretization using 128\,$\times$\,32\,$\times\,$1 FD cells and the RK23 time stepping scheme with a time step around 600\,fs (dynamically adapted during the simulation) was used.  The relatively large time steps used to solve this standard problem demonstrate that \mumax incorporates robust time stepping schemes with accurate adaptive step mechanisms. The adaptive step algorithms ensure optimal time step lengths and thus reduce the number of field evaluations, speeding up the simulation. Here, a total time of only 2.5 seconds was needed to finish this simulation on the GPU compared to 16 seconds on the CPU. In this case the speedup on GPU is limited due to the small number of FD cells. When the simulation is repeated with a finer discretization of 256\,$\times$64\,$\times$\,2 cells, on the other hand, the GPU speedup already becomes more pronounced: the simulation finishes in 46 seconds on the GPU but takes 20'32" on the CPU.

From Figs. \ref{stdprobl4_1} and \ref{stdprobl4_2} it is clear that the results obtained with \mumax are well within the spread of the curves obtained by other authors.  Also the magnetization plots are in close agreement with those available at the {$\mu$}Mag website \cite{mumag}.

\subsection{Spin-transfer torque standard problem \cite{Najafi2009}}
\begin{figure}
\begin{center}
\includegraphics[width = \columnwidth]{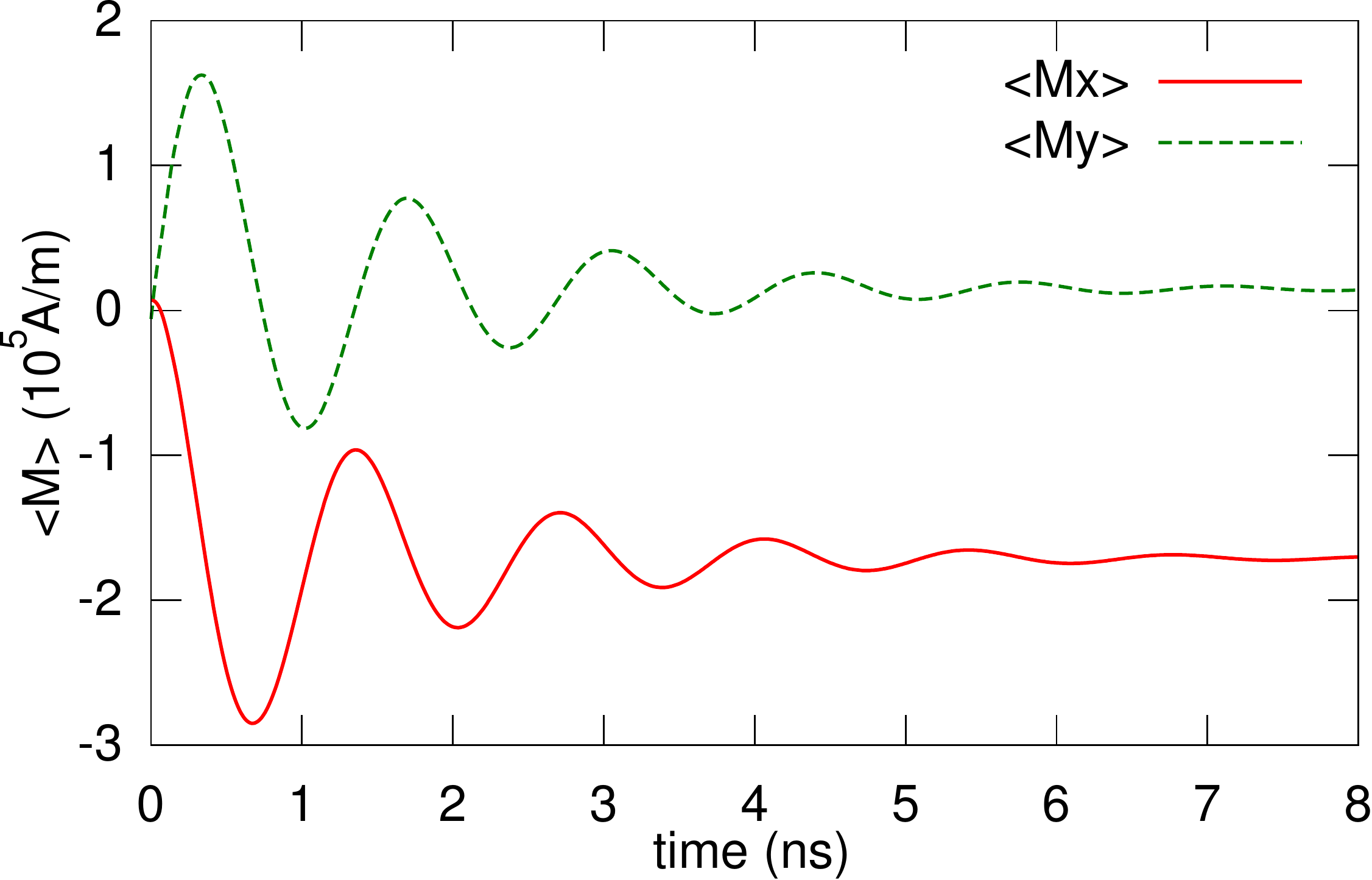}
\end{center}
\caption{Time evolution of the average in plane magnetization during the first 8 ns of the spin-transfer torque standard problem.  To facilitate the visual comparison of our results with \cite{Najafi2009}, the average magnetization is expressed in [A/m] and the same axes ratios are chosen. \label{stdprobl5_1}}
\end{figure}

\begin{figure}
\begin{center}
\includegraphics[width = 0.7\columnwidth]{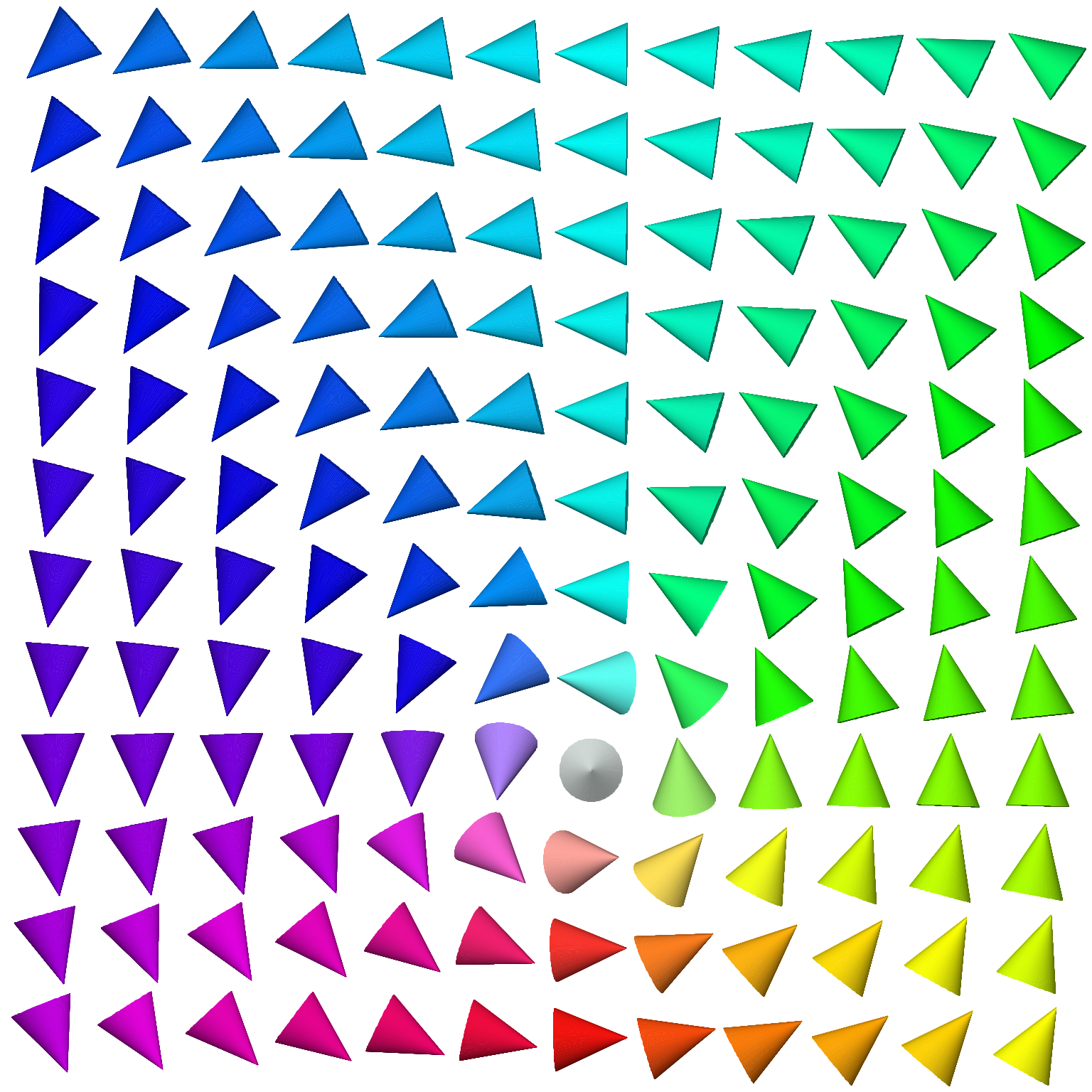}
\end{center}
\caption{Magnetization configuration at t=0.73\,ns as found during the simulation of a standard problem incorporating spin-transfer torque \cite{Najafi2009}.  The vortex core evolves towards a new equilibrium state under influence of a spin-polarized dc current allong the horizontal direction.This figure is rendered with the built-in graphics features present in \mumax. \label{stdprobl5_2}}
\end{figure}

$\mu$Mag does not propose any standard problems that include spin-transfer torque.  Therefore we rely on a standard problem proposed by M. Najafi et al. \cite{Najafi2009} to check the validity of the spin-transfer torque description implemented in \mumax.  The standard problem considers a permalloy sample ($A=1.3\times10^{-11}$\,J/m,  $M_s=8.0\times10^5$\,Am$^{-1}$) with dimensions 100\,nm\,$\times$\,100\,nm\,$\times$\,10\,nm.  The initial equilibrium magnetization state is a predefined vortex, positioned in the center of the sample and relaxed without any spin-transfer torque interaction ($\alpha=1.0$).  Once relaxed, an homogeneous spin-polarized dc current  $\mathbf{j}=10^{12}$\,Am$^{-2}$ along the $x$-axis is applied on the sample.  Now, $\alpha$ is $0.1$ and the degree of non-adiabicity $\xi$ is 0.05, see expression (\ref{STT}).  Under these circumstances, the vortex center moves towards a new equilibrium position.  The time evolution of the average in plane magnetization and the magnetization configuration at $t$=0.73\,ns are shown respectively in Fig. \ref{stdprobl5_1} and Fig. \ref{stdprobl5_2}.  The results are in good agreement with those presented in reference \cite{Najafi2009}. With a discretization of 128\,$\times$\,128 FD cells, 10 minutes of simulation time were needed to obtain the presented data.

\section{Performance}
\begin{figure}
\begin{center}
\includegraphics[width = \columnwidth]{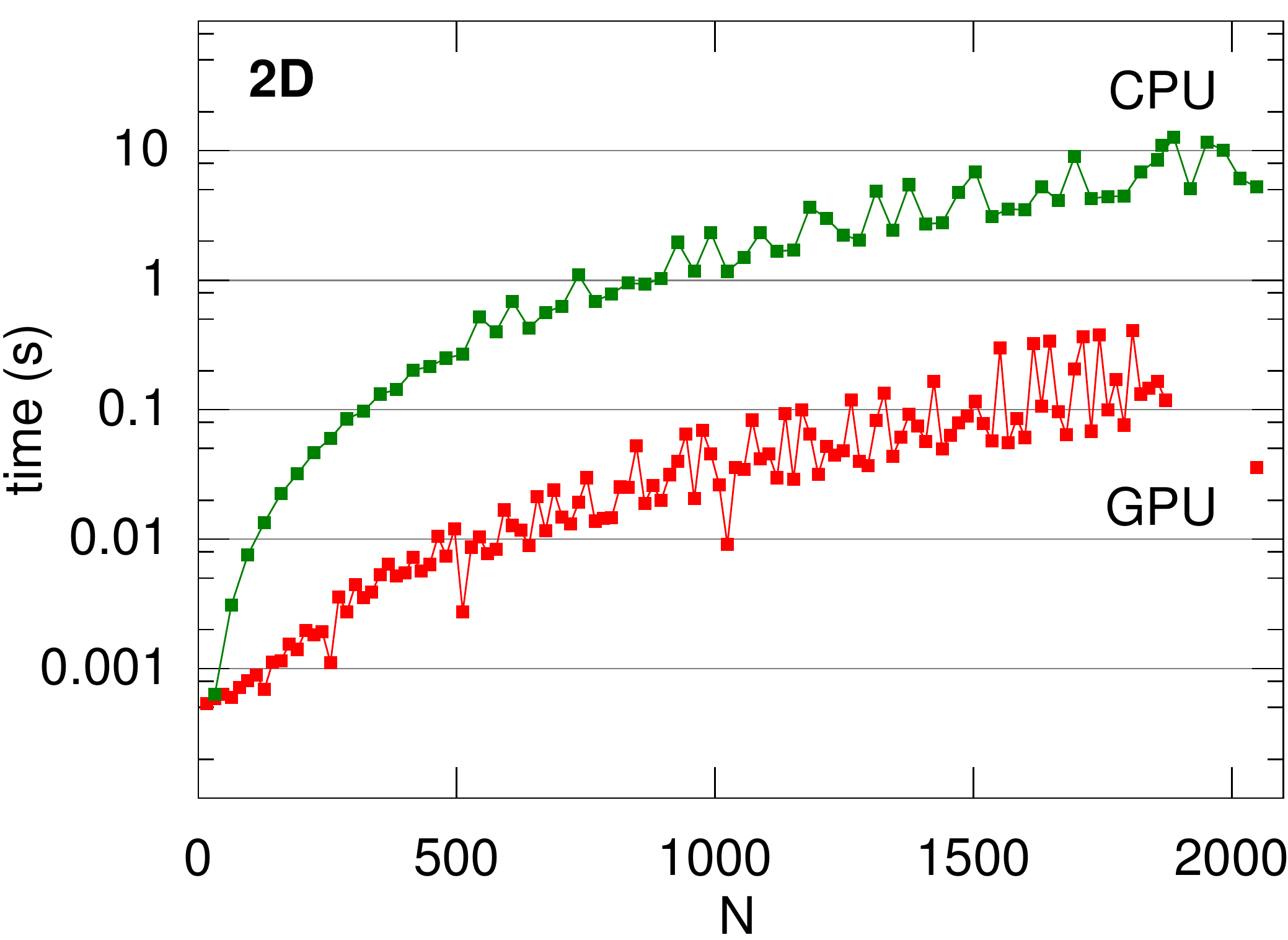}\\ \vspace{0.5cm}
\includegraphics[width = \columnwidth]{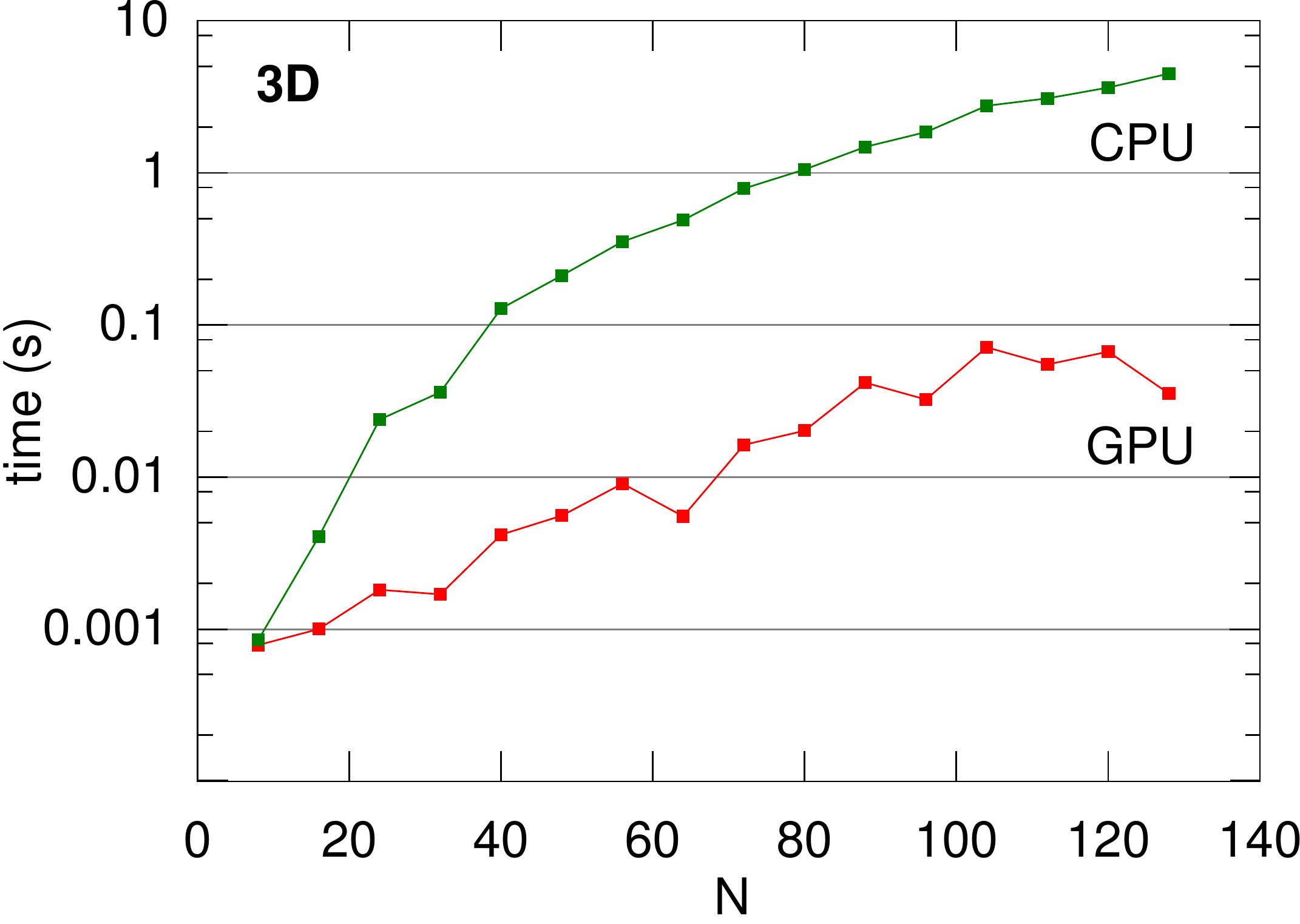}
\end{center}
\caption{Time required to perform one time step using the Euler method for (top) 2D geometries with varying dimensions N$\times$N and (bottom) 3D geometries with varying dimensions N$\times$N$\times$N.  The CPU computations are performed on a 2.8\,GHz intel core i7-930 processor, while the GPU computations are performed on nVIDIA GTX580 GPU hardware. \label{timing}}
\end{figure}

The performance of \mumax on the CPU is roughly comparable to OOMMF. The CPU performance is thus good, but our main focus is optimizing the GPU code. Special attention went to fully exploiting the numerical power of the GPU while focussing on the time- and memory-efficient simulation of large micromagnetic problems.  Figure \ref{timing} shows the time required to take one time step with the Euler method (i.e. effective field evaluation, evaluation of the LL-equation and magnetization update) on CPU (1 core) and on GPU for 2D and 3D simulations.  

In both the 2D and 3D case, speedups of up two orders of magnitude are obtained for large dimensions.  For smaller geometries, the speedups decrease but remain significant.  This can be understood by the fact that in these simulations not enough FD cells are considered to have all $\mathcal{O}(10^4)$ available threads at work at the same time.  Hence, the computational power is not fully exploited.  Furthermore, Fig. \ref{timing} shows that the CPU performance as well as the GPU performance does not follow a smooth curve.  This is a consequence of the FFTs which are most efficient for powers of two, possibly multiplied with one small prime (in the benchmarks shown in Fig. \ref{timing}, the default rounding to these optimal sizes is not performed).  In the GPU implementation this is even more the case than in the CPU implementation. E.g., the 2D simulation with dimensions 992\,$\times$\,992 is five times slower than the 2D simulation with dimensions 1024$\times$1024.  This shows that not only for accuracy reasons, but also for time efficiency reasons, it is most advantageous to restrict the simulations domain to the optimal dimensions defined by $2^n \times \{1,3,5 \mathrm{\ or\ } 7\}$.  Because of the extreme impact on the performance of \mumax, we opted to standardly rescale the size of the FD cells such that the dimensions are rounded of to one of these optimal sizes.

\begin{table}
\caption{Time needed to take one time step with the Euler method on CPU and GPU for 2D geometries (top) and 3D geometries (bottom). \label{table}}
\begin{center}
\begin{tabular}{|l|l|l|l|}
\hline
%
size & CPU time (ms) & GPU time (ms) & speedup \\\hline
$32^2$ & 0.633 & 0.59 & $\times$ 1.07\\
$64^2$ & 3.092 & 0.60 & $\times$ 5.1 \\
$128^2$ & 6.739 & 0.69 & $\times$ 9.7\\
$256^2$ & 59.90 & 1.11 & $\times$ 18\\
$512^2$ & 266.8 & 2.75 & $\times$ 47\\
$1024^2$ & 1166 & 9.07 & $\times$ 128\\
$2048^2$ & 5233 & 35.78 & $\times$ 146\\\hline
$8^3$ & 0.8492 & 0.79 & $\times$ 1.07\\
$16^3$ & 4.066 & 1.03 & $\times$ 3.9\\
$32^3$ & 36.14 & 1.70 & $\times$ 21\\
$64^3$ & 489.6 & 5.52 & $\times$ 88\\
$128^3$ & 4487 & 35.42 & $\times$ 126\\\hline
\end{tabular}
\end{center}
\end{table}

\begin{figure}
\begin{center}
\includegraphics[width = \columnwidth]{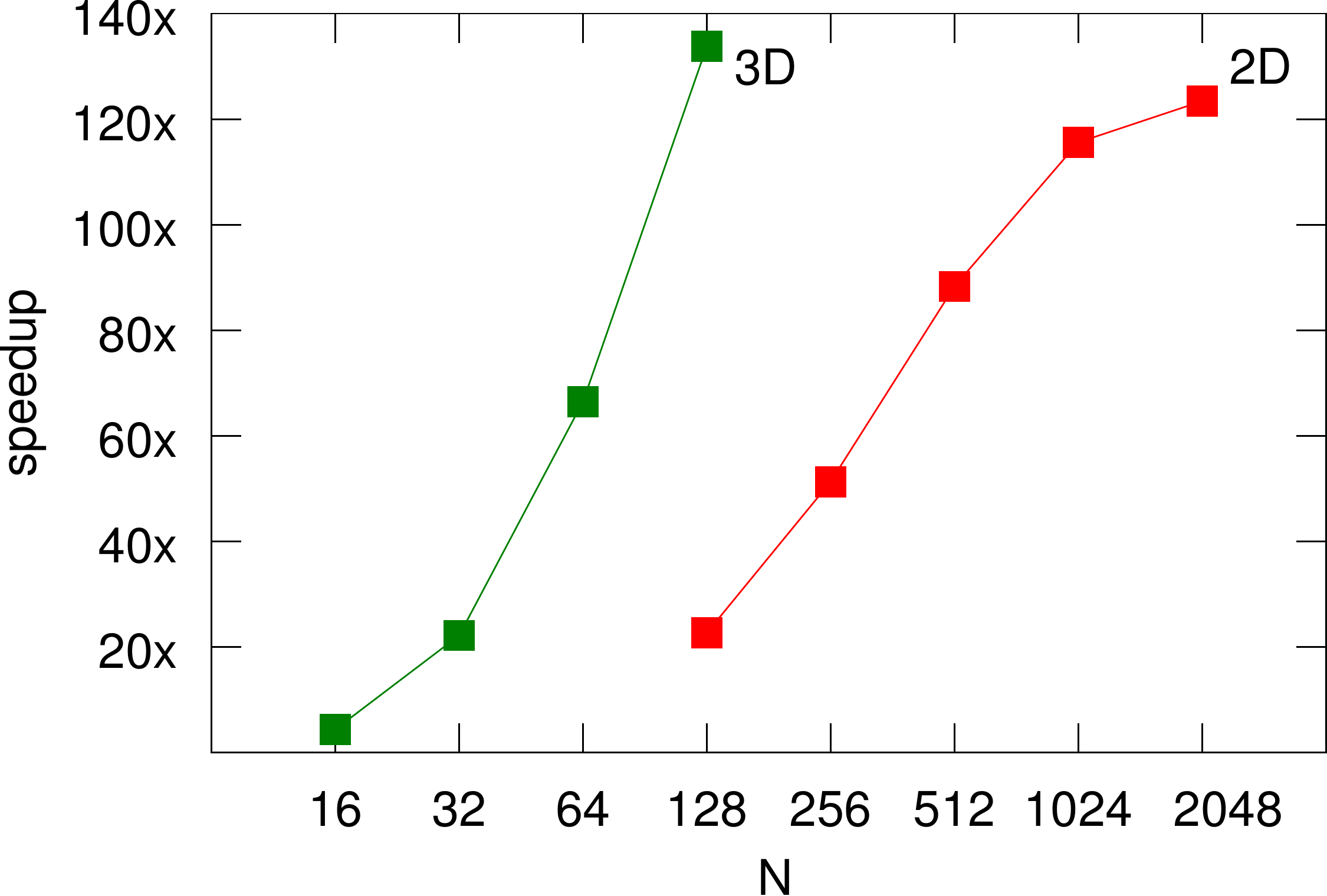}
\end{center}
\caption{Speedup obtained with \mumax running on a GTX580 GPU compared to OOMMF on a 2.8GHz core i7-930 CPU.  The 2D and 3D geometries have sizes N$\times$N and N$\times$N$\times$N respectively.  The lowest speedup for the 16 x 16 x 16 case -- an unusually small simulation-- is still a factor 4. \label{performance}}
\end{figure}

Due to the typical architecture of GPUs and the nature of the FFT algorithm, simulations of geometries with power of two sizes run extremely fast on GPU.  Table \ref{table} gives an overview of the speedups for these sizes for the 2D and 3D case and Fig. \ref{performance} shows the speedup obtained for these power of two sizes by \mumax compared to the OOMMF code.  Both comparisons result in speedups larger than a factor 100.  This means that simulations that used to take several hours can now be performed in minutes.  The comparison between the speedups shown in Table \ref{table} and Fig. \ref{performance} further show that our CPU implementation has indeed a comparable efficiency regarding to OOMMF.  The immense speedups evidence the fact that \mumax can indeed open completely new research opportunities in micromagnetic modelling.

\section{How to use \mumax}

\mumax is released as open source software under the GNU General Public License (GPL) v.3 and can thus be freely used by the community. In addition to the terms of the GPL, we kindly ask to acknowledge the authors in any publication or derivative software that uses \mumax, by citing this paper.  The \mumax source code can be obtained via \texttt{http://dynamat.ugent.be/mumax}. To use the software, a PC with a "CUDA capable" nVIDIA GPU and a recent 64-bit Linux installation is required. 

A \mumax simulation is entirely specified by an input file passed via the command-line. I.e., once the input file is written, no further user interaction is necessary to complete the simulation. This allows, for instance, to run large batches of simulations unattended. Nevertheless, the progress of a simulation can easily be checked: the number of time steps taken, total simulated time, etc. is reported in the terminal, PNG images of the magnetization state can be output on-the-fly, graphs of the average magnetization can easily be obtained while the simulation is running, etc. Furthermore, \mumax's output format is compatible with OOMMF, enabling the use of existing post-processing tools to visualize and analyze the output. Built-in tools for output processing are available as well. The 3D vector field in Fig. \ref{stdprobl5_2}, e.g., is rendered by \mumax's tools.

\mumax input files can be written in Python. This offers powerful control over the simulation flow and output. As an example, the code snippet below simulates an MRAM element as in standard problem 2 (see section \ref{prob2}). Starting form a uniform state in the $+x$ direction, it scans the field $B$ in small steps until the point of coercivity. This illustrates how easily complex simulations can be defined.

\begin{figure}
\small
\begin{verbatim}
from mumax import *

msat(800e3)
aexch(1.3e-11)
partsize(500e-9, 50e-9, 5e-9)
uniform(1, 0, 0)

B = 0
while avg_m('x') > 0:
      staticfield(-B, 0, 0)
      relax()
      B += 1e-4
# B now holds the coercitive field
save('m', 'binary')
\end{verbatim}
\normalsize 
\caption{\mumax input file snippet illustrating a simulation specification in Python. After initialization, a field in the $-x$ direction is stepped until the average magnetization along $x$ reaches zero. The possibility of writing conditional statements and loops, and obtaining information like the magnetization state allows to construct arbitrarily complex simulation flows.}
\end{figure}

\section{Conclusions and Outlook}

\mumax is the first GPU-based micromagnetic solver that is publicly available as open-source software.  Due to the large number of considered interaction terms and the versatile geometrical options (e.g. periodic boundary conditions) the software covers many of the classical micromagnetic research topics.  The code is extensively validated by considering several standard problems and is shown to be reliable.  The time gains are extremely large compared to CPU simulations: for large simulations a speedup with a factor 100 is easily obtained.  These enormous speedups will open up new opportunities in micromagnetic modelling and boost fundamental magnetic research.

In the future, \mumax will be extended towards a yet more multipurpose software package incorporating other interactions in the Landau-Lifshitz equation: other expressions for the exchange contribution, exchange bias, magnetoelastic coupling, etc.  Boundary correction methods to help in approximation non-square geometries better are currently also being considered. Furthermore, the description of more complex, non uniform microstructures will be made possible as e.g. nanocrystalline materials. Modules for hysteresis research and magnetic domain studies will be developed following the presented 2D and 3D approach as well as the so-called 2.5D approach (infinitely thick geometries).  

Furthermore, the efficient simulation of yet larger problems is planned by introducing multiple GPUs --in one or more machines-- for one single simulation.  This way, the up to now limited memory available on the GPU hardware can be circumvented and the computational power will be further increased.  Apart from requiring efficient communication between the different GPUs, this should be possible without drastic changes to our code as most of the implementation is already hardware-independent.

\section*{Acknowledgements}
Financial support from the Flanders Research Foundation (FWO) is gratefully acknowledged.  We cordially thank Bartel Van Waeyenberge, Luc Dupré, and Dani\"el De Zutter for supporting this research.  Furthermore we would also like to thank André Drews, Claas Abert, Gunnar Selke and Theo Gerhardt from Hamburg University for the fruitful discussions and feedback.

\section*{References}
\bibliography{bibliography}

\end{document}